\documentclass[aps,prb,twocolumn,groupedaddress,floatfix,showpacs]{revtex4-2}
\usepackage{graphicx}
\usepackage{amsmath}
\usepackage{subcaption}
\usepackage{color}
\definecolor{green}{rgb}{0,0.5,0}

\newcommand{\parl}{\parallel}
\newcommand{\beq}{\begin{equation}}
\newcommand{\bearr}{\begin{eqnarray}}
\newcommand{\eeq}{\end{equation}}
\newcommand{\earr}{\end{eqnarray}}

\begin{document}

%\preprint{}

\title{Rydberg excitons in core-shell nanostructures \\ I. Optical properties}


%\thanks{}
\author{Sylwia Zieli\'{n}ska-Raczy\'{n}ska}
\author{Gerard Czajkowski}
\author{David Ziemkiewicz}
\email{david.ziemkiewicz@pbs.edu.pl}
 \affiliation{Department of
 Physics, Technical University of Bydgoszcz,
\\ Al. Prof. S. Kaliskiego 7, 85-789 Bydgoszcz, Poland}


\date{\today}


%\keywords{}


\begin{abstract}
Optical properties of Cu$_2$O-metal core-shell nanostructures containing Rydberg excitons are examined. Calculations of electron and hole eigenfunctions and eigenenergies are used to calculate exciton energy and its dependence on system geometry and dimensionality. The limits of very small and very large structures are discussed, making the results applicable to low principal quantum number excitons as well as to Rydberg states. The Real Density Matrix Approach  method is used to calculate the optical properties of the studied systems.
\end{abstract}


%\keywords{}

\maketitle

\section{Introduction}

Excitons, elementary excitations in semiconductor consist of a Coulomb-bound pair of an electron and a hole.  In a recent high-resolution laser absorption study \cite{Kazimierczuk} it was  shown that Cu$_2$O hosts  excitons up to $n=30$ which opened the portal to the field of giant Rydberg excitons in a solid state which is an analogon of highly excited Rydberg atoms in atomic physics. The electron-hole pair is weakly bounded and may extend over many thousands of lattice unit cells, thus its interaction is screened by  the static permittivity of the semiconductor. Rydberg excitons (RE) have remarkable features; they have giant microscopic dimension (up to 1$\mu$m), long lifetimes reaching nanoseconds \cite{Heck017, Schweiner2017, Thomas, Ertl, Kim} and they are very sensitive to external fields because of their huge polarizibility. In recent decade  there has been growing interest in examining spectroscopic and optical properties of Rydberg excitons in a natural and artificial Cu$_2$O bulk system \cite{Heck2017, Steinhauer, Morin2022, Sylwiaqb}. 

Due to their unusual properties, Rydberg excitons are considered major candidates towards solid-state quantum systems which provide unprecedented capabilities for the realization of novel devices owing to their robustness, scalability \cite{Heck2017, Kim} and miniaturization ability \cite{Neubauer}. 
Nanoparticles \cite{Hamid} and other low dimensional systems \cite{Qdisk, Belov2024, Konzel2020} are natural and suitable system for quantum confinement and interesting platform for realizing quantum technologies with Rydberg excitons. 

Takahata \emph{et al.} \cite{Takahata2018} performed observations of non-local response of weakly confined RE in plane-parallel Cu$_2$O films, Konzelmann \emph{et al}\citep{Konzel2020} studied theoretically the influence of quantum confinement effect resonance shifts, Orfanakis \emph{et al} \cite{Hamid} observed broadened Rydberg excitons transitions in nanoparticles confirming experimentally theoretical predictions for quantum dots, wires and wells with RE presented by Ziemkiewicz  \emph{et al} \citep{Qdisk}.

In the last couple of years, there has been extensive progress in the synthesis of core-shell nanomaterials of various shapes which  has led to the  their exploration for use in plasmonics, sensing, optolectronics and in photochemistry \cite{Sharma,Bartolucci2024,Meir,Wang}. Their crucial advantage is tunability of physical properties by modifying the shell composition and thickness. Due to this flexibility, core-shell nanomaterials turned out to be highly beneficial in optical and electronic applications, where the core-shell architecture enables fine control over band-gap and resonance engineering \cite{Wang}. Particularly, hybrid nanostructuters consisting of a metal core surrounded by semiconductor layer are promising candidates for applications in optoelectronics, electro-optics and solar engineering because  they may retain the properties of each component as well as exhibit new advantages due to combination of both materials.

Physics of excitons has taken the advantage of these multi-component nanostructures
 \cite{Rana2021, Hibibi}, with their unique structural configurations allowing for precise control over quantum confinement effects of excitons inside. One of the most significant optical features of core-shell nanomaterials is the ability to tune their absorption and emission spectra through confinement effects, leading to improvements in optical, electrical and magnetic characteristic of such nanostructures. Therefore, it seems to be interesting to examine Rydberg excitons behavior in such type of nanostructures.

In the presented paper, we extend our previous studies of confined Rydberg excitons in low dimensional systems \cite{Qdisk, Hamid} by considering multi-component core-shell nanostructures. In particular, we analyze a metallic core material surrounded by a Cu$_2$O shell, where an exciton is confined. This confinement has the potential to drastically alter the shell's optical properties; the thickness of  the shell influences the exciton's binding energy and resonance positions \cite{Hibibi,Nandan}. The presence of a metallic core opens up possibility of plasmon-exciton studies \cite{Antos2014,Stete}; with the recent advancements in Cu$_2$O core-shell structure fabrication \cite{Bartolucci2024,Meir}, a theoretical description of Rydberg excitons in such a system is needed.

We adopt the  method known as the Real Density Matrix Approach (RDMA) to calculate the optical functions for Rydberg excitons with arbitrary high principal quantum numbers in core-shell systems of a  various geometry with reduced dimensionality, examining the impact of dimensionality and symmetry on the system optical properties. RDMA calculations are supplemented by \emph{ab initio} numerical calculations of the exciton state wavefunctions and eigenenergies.

The paper is organized as follows. In the first section, the basic equations of RDMA are presented and general calculation assumptions are outlined. Then, three subsections are dedicated to individual considered geometries: 3-dim spherical core-shell particles and  2-dim quantum rings  and  short outline devoted to an intemediate, regarding to dimensionality, system of quantum rods. Next section is devoted to the calculation results. Again, individual subsections are dedicated to specific geometries and a comparison between them. Next, conclusions are outlined. The paper ends with two appendices dedicated to the modeling of the Coulomb potential in the presented approach and numerically calculating the exciton wavefunctions and eigenenergies.

\section{Theoretical description}
We use the real density matrix approach to describe the linear optical properties of Cu$_2$O Core - Shell nanostructures, with the emphasis on Rydberg excitons. In this approach, the optical properties are described by a system of 3 equations:
\begin{enumerate}
\item The so-called constitutive (material) equation, \item
Maxwell's field equation, \item The equation defining the
polarization. \end{enumerate} We use the constitutive equation in
the form \cite{Dima}
\begin{equation}\label{1}
-i(\hbar \partial_t + \Gamma) Y_{eh}+H_{eh}Y_{12} ={\bf M}{\bf E},
\end{equation}
where  $Y_{eh}$ is the so-called coherent amplitude (exciton
amplitude) being function of electron-hole pair  coordinates ${\bf r}_h$ and ${\bf r}_e$ , $H_{eh}$ is the two-band Hamiltonian  with gap $E_{g}$, and $\Gamma$ is a
phenomenological damping coefficient. In the above equation ${\bf
M}(\textbf{r})$ is a smeared-out transition dipole density
%depending on the coherence radius $r_0 =\left(E_g/2\mu\hbar^2\right)^{-1/2}$; the $E_g$ is the fundamental gap and $\mu$ is reduced effective mass of the electron-hole pair.
and \textbf{r} is a relative electron-hole distance.

The system is irradiated by a normally incident electromagnetic wave of a frequency $\omega$, a wave vector $k$, with amplitude $\varepsilon$ and polarization $\textbf{e}$
\begin{equation}\label{eq:field}
\textbf{E}(z, t)= \textbf{e}\varepsilon(z,t)\exp[i(\omega t-k z )].
\end{equation}

Here, we consider a two-band effective mass Hamiltonian $H_{eh}$, which includes the electron and hole kinetic energy, the electron-hole interaction potential and the confinement potentials. In
consequence, the two-band Hamiltonian $H_{eh}$ with the gap energy $E_g$  for any pair of bands is
\begin{eqnarray} \label{hamilt}
&&H_{eh}=E_{g} +\frac{{\bf
p}_{h}^2}{2m_{h}}+\frac{{\bf
p}_{e}^2}{2m_{e}}\\&& +V_{c} +V_{conf,e}+V_{conf, h},\nonumber
\end{eqnarray}\\
\noindent where the second and the third term on the r.h.s. give
the hole and the electron kinetic energy operators with band
effective masses, the  fourth term represents the electron-hole
attraction and the two last terms are the surface confinement
potentials for the electron and hole. The total polarization of the
medium   is related to the coherent amplitudes by
\begin{equation}\label{Polar}
{\bf P}({\bf R})=2 \hbox{Re}\int d^3{r}{\bf
M}({\bf r}) Y_{eh}({\bf R},{\bf r}),
\end{equation}
\noindent where  $\textbf{R}$ is the center-of mass coordinate $$\textbf{R}=\frac {m_h \textbf{r}_1+m_e\textbf{r}_2}{m_h+m_e}$$   The above equations
 form the set of constitutive equations which
connect linear polarization to the electric field, which in
addition must obey Maxwell's equations
\begin{equation}\label{Maxwell}
c^2\nabla^2 {\bf E(R)} - \epsilon_b \ddot{\bf E} =
\frac{1}{\epsilon_0}{\bf \ddot{P}(R)}.
\end{equation}
where $\epsilon_b$ is the dielectric permittivity of the medium.
Using the long wave approximation we obtain the coherent amplitude
$Y_{12}$ from Eq. (\ref{1}) then, from Eq. (\ref{Polar}) one can
determine the excitonic susceptibility
$\chi(\omega,\textbf{k})$, which is given by
\begin{equation}\label{Polar2}
\textbf{P}(\omega, \textbf{k})=\epsilon_0\chi(\omega,\textbf{k})\textbf{E}(\omega,\textbf{k}),
\end{equation}
where $\varepsilon_0$ is the vacuum dielectric constant. 
\\
In this paper, we focus on two types of core-shell nanostructures in the order of reducing dimensionality: spherical particles (3 dimensions) and  quantum rings (2 dimensions) with particular case of a quantum rod, which is an example of quasi 2 dimensional system and is an intermediate type between a core-shell and a quantum ring. 

In all cases, we follow a set of simplifying assumptions:
\begin{enumerate}
\item The structure consists of a central metal core characterized by radius $r_1$ surrounded by a shell made out of Cu$_2$O, characterized by a radius $r_2$. Structure is surrounded by air and the two medium interfaces form surfaces with geometry-dependent shape.
\item Both electron and hole are confined to the shell between the two surfaces, with infinite potential barriers at the metal-Cu$_2$O and Cu$_2$O-air interface. 
In addition to the confinement, the carriers (electron and hole) are interacting via dielectrically screened Coulomb potential $V_c$.
\item The unique property of  Cu$_2$O is that the effective mass of the electron in  is much greater than the mass of the hole;  so that the hole moves in a field of a centrally symmetric attracting potential, 'smeared out' by the electron.
\item Electron is smeared out on the metal-Cu$_2$O interface.
\end{enumerate}
In order to obtain exciton state wavefunctions and eigenenergies   one has to solve the Schr\"{o}dinger equation apropriate for a system of a particular geometry. \\
 As mentioned above, the electron is in average immobile and confined by the  potential; its eigenfunctions are given by the Schr\"{o}dinger equation 
\begin{eqnarray}
\Biggl[-\frac{\hbar^2}{2m_e}\hbox{\boldmath$\nabla$}^2_{r_e}+V_{conf}(r_e)\Biggr]\psi(\textbf{r}_e)=E_e\psi(\textbf{r}_e),
\end{eqnarray} 
  and the Schr\"{o}dinger equation for the hole is the following
\begin{eqnarray}
\Biggl[-\frac{\hbar^2}{2m_h}\hbox{\boldmath$\nabla$}_{r_h}^2+V_c+V_{conf}(r_h)\Biggr]\psi(\textbf{r}_h)=E_h\psi(\textbf{r}_h).
\end{eqnarray}
The form of the Laplacian and the confinement potentials for both particles depend on the geometry and dimensionality of the system; the total energy of an exciton is a sum of eigenenergies of an electron and a hole.

After finding the eigenenergies of individual excitonic states, one can proceed to calculate optical functions of the medium with the help of RDMA. In particular, using Eq. (\ref{1}) in stationary limit and Eqs (\ref{hamilt},\ref{Polar},\ref{Polar2}) one obtains the following formula for the excitonic susceptibility 
\begin{equation}\label{susc}
\chi=\frac{2 M_0^2}{\epsilon_0{\mathcal E}}\sum_{n_e,n_h}\frac{\vert \textbf{M}\cdot
 \textbf{E}\vert^2 f}{E_g-\hbar\omega-i\Gamma +E_{n_e}+E_{n_h}},
\end{equation}
which imaginary part allows to get an absorption spectrum. In general, the summation is done over quantum numbers $n,l$, but here we focus only on a specific case of $P$ excitons characterized with $l=1$. The energies $E_{ne}$ and $E_{nh}$ are the electron and hole eigenenergies,
\begin{equation}\label{oscstrg}
f\cdot M_0^2=\left[\int\limits_{\rho_1}^{\rho_2} \rho^\gamma d\rho M(\rho)\psi(\rho)\right]^2,
\end{equation}
is the effective oscillator strength. It is dependent on the dimensionality of the system; $\gamma=2$ (3 dim) and $\gamma=1$ (2 dim);  $\rho=r/a^*$ is the reduced radius, $a^* \approx 1.1$ nm is the Bohr radius, and the smeared-out dipole density $M(\rho)$ is taken in the form
\begin{equation}\label{dipoledensity} 
M(\rho)=N_\gamma M_0\,\exp\left(-\frac{\rho}{\rho_0}\right),
\end{equation}
where $\rho_0$ is the so-called coherence radius, 
\begin{equation}\label{rho_0}
\rho_0=\sqrt{\frac{R^*}{E_g}},
\end{equation}
$M_0$ is the dipole matrix element, and 
\begin{equation}\label{engamma}
N_\gamma=\left(\int\limits_{\rho_1}^{\rho_2} \rho^\gamma\,d\rho\,
\exp\left(-\frac{\rho}{\rho_0}\right)\right)^{-1}.
\end{equation}
The quantity from the numerator of Eq.(\ref{susc})
\begin{equation}
\vert \textbf{M}\cdot\textbf{E}\vert={\mathcal E}\int\limits_{\rho_1}^{\rho_2}\rho^2\,d\rho \psi(\rho),
\end{equation}
follows from Eq. (\ref{oscstrg}) and 
%and $\textbf{r}_e$ is treating as the exciton center-of-mass coordinate. 
$\mathcal E$ is the propagating electromagnetic wave amplitude, as defined in Eq. (\ref{eq:field}).\\\\
Summarizing the presented calculation scheme - first the Schr\"{o}dinger equations   for an electron and a hole for the particular system geometry and dimensionality  are solved to obtain exciton wavefunctions and eigen-energies, then  the excitonic amplitude $Y_{12}$, obtained from Eq. (\ref{1})  is inserted into Eq. (\ref{Polar}) giving the polarization and finally, the susceptibility Eq. (\ref{susc}) from which all optical functions of a considered system is calculated.  We start from the most general case of 3-dimensional system and then proceed to the more confined geometries.
\subsection{Core-shell particle}
\begin{figure}[ht!]
\centering
\includegraphics[width=.8\linewidth]{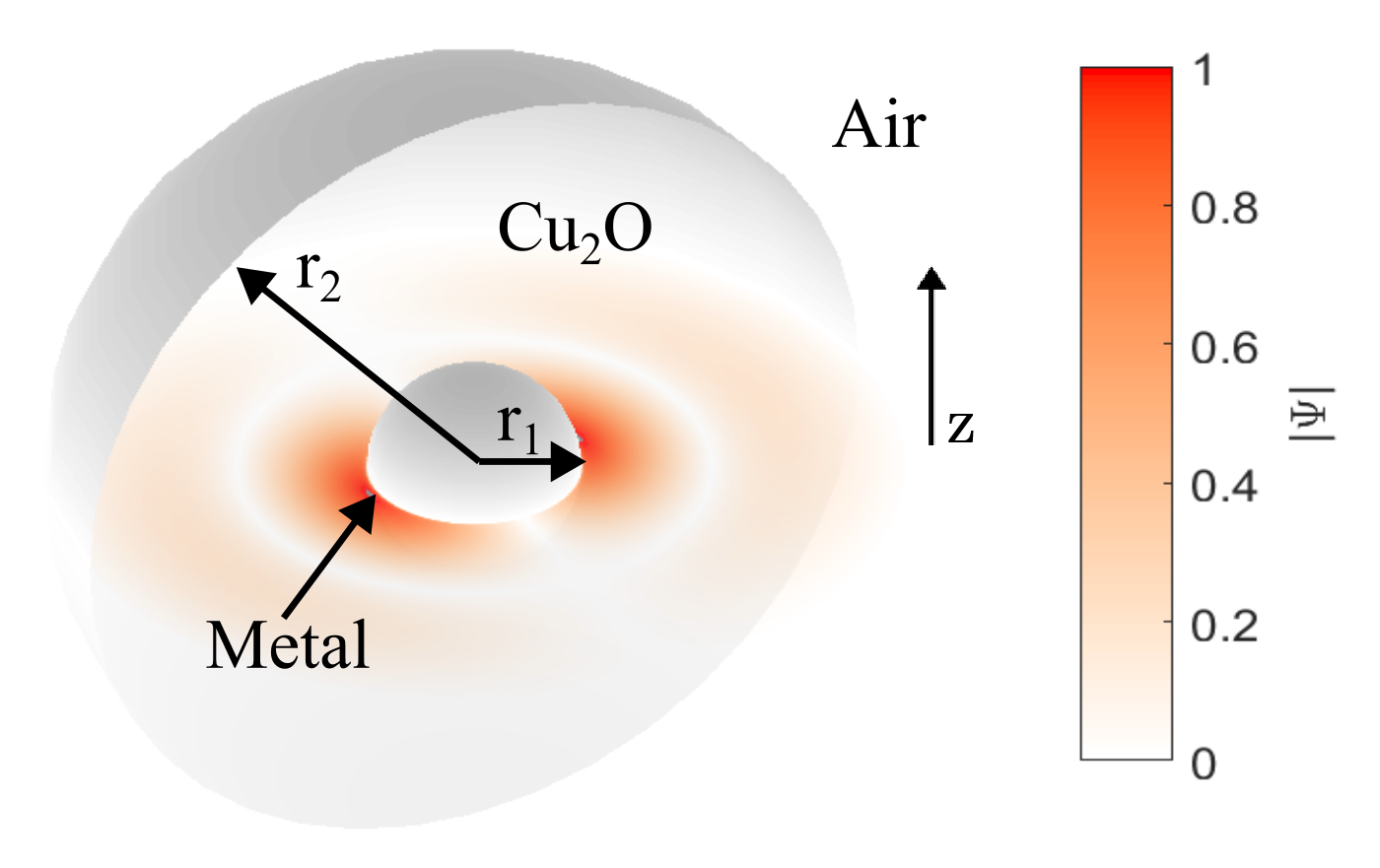}
\caption{Schematic of the considered system; metal ball with a radius $r_1$ is surrounded by a layer of Cu$_2$O with radius $r_2$. Color marks a cross-section of the hole wavefunction of $2P$ exciton.}\label{fig:1}
\end{figure}

We consider a layered sphere with core radius $r_1$, and a shell of thickness
$h=r_2-r_1$, as shown on Fig. \ref{fig:1}. The layer is made out of Cu$_2$O and is characterized by permittivity $\epsilon_b=7.37$. Both the metal core and the air outside act as an infinite potential barriers. 
The Schr\"{o}dinger equations for the an electron and a hole follow from Eqs(7-8) and have the following forms
\begin{eqnarray}\label{shrod_sph}
&&\Biggl[\frac{-\hbar^2}{2m_e}\hbox{\boldmath$\nabla$}^{(3D)2}_{r_e}+V_{conf}(r_e)\Biggr]\psi(\textbf{r}_e)=E_e\psi(\textbf{r}_e),\\
&&\left[\frac{-\hbar^2}{2m_h}\hbox{\boldmath$\nabla$}^{(3D)2}_{r_h}+V_c+V_{conf}(r_h)\right]\psi(\textbf{r}_h)=E_h\psi(\textbf{r}_h).
\end{eqnarray}
It is assumed that the Coulomb potential has generalized form
\begin{equation}\label{Coulomb}
V_c=\frac{e^2}{4\pi\epsilon_0\epsilon_b (r-\alpha r_1)},
\end{equation}
with a parameter $\alpha$ ($0 \leq \alpha \leq \frac{r_2}{r_1})$. Such a form of screening potential enables one to preserve a system symmetry and in general allows for maintaining continuity between solutions in various regimes, see Appendix A for details). The confinement potential is
\begin{eqnarray}\label{boundarysphere}
&&V_{conf}(r_{e,h})=\left\{ \begin{array}{ll}
0\quad \mbox{for}\quad r_1\leq\quad r_{e,h}\leq r_2,\\
\infty\quad \mbox{for}\quad r_{e,h}>r_2\quad\hbox{or}\; r_{e,h}<r_1.
\end{array}\right.
\end{eqnarray}
\\
Such a form of the potential in the spherical core-shell geometry leads to confine  an electron in a  sperical quantum well, therefore eigenfunctions of Eq. (\ref{shrod_sph})  are the spherical Bessel functions of the first kind \cite{Abramowitz}
\begin{eqnarray}
&& w_{\ell m n}(r_e,\theta_e,\phi_e)=C_n Y_{\ell m}(\theta_e,\phi_e)\\
&&\times\frac{1}{k_n^2 (r_e-r_1)}\sin k_n(r_e-r_1),\nonumber\\
&&k_N=\frac{N\pi}{r_2-r_1},\nonumber
\end{eqnarray}
 $\ell=0,1...$, $m=-\ell,...,\ell$ are  the orbital   and  magnetic quantum numbers, respectively and $Y_{lm}(\theta_e,\phi_e)$ are spherical harmonics.
Eigenenergies are given by
\begin{equation}
E^{e}_{n}=\left(\frac{n\pi}{r_2-r_1}\right)^2 R^*_e.
\end{equation}
\\
The solution  of Eq.(16) for the holes is a modification of a hydrogenlike wavefunction that fulfills a structure-specific boundary conditions
\begin{equation}\label{falowa_sph}
\psi(\textbf{r}_h)=Y_{\ell m}(\theta_h,\phi_h) R(r),
\end{equation}
with radial part $R(r)$, which fulfills the equation
\begin{equation}\label{rad_1}
\frac{\partial^2 R}{\partial r^2}+\frac{2m_h}{\hbar^2}\left[E+V_c-\frac{\ell(\ell+1)}{r^2}+V_{conf}(r)\right]R=0,
\end{equation}  
with quantum number $\ell=0,1...$.A numerical solution of the above equation, as described in Appendix A, yields eigenenergies; radial wavefunctions combined with spherical harmonics provide complete wavefunctions, as indicated in Eq. (\ref{falowa_sph}).

\subsection{Quantum ring}
\begin{figure}[ht!]
\centering
\includegraphics[width=.8\linewidth]{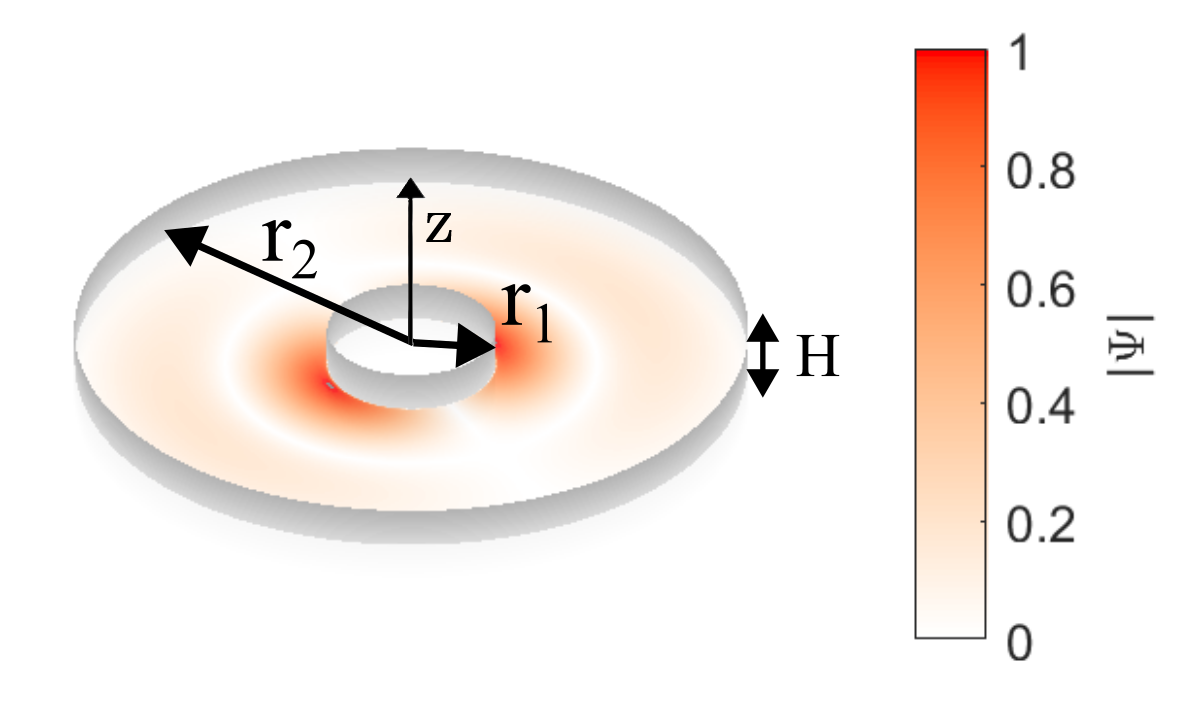}
\caption{Schematic of the quantum ring system; metal cylinder with a radius $r_1$ is surrounded by a layer of Cu$_2$O with radius $r_2$. Color marks a cross-section of the hole wavefunction of $2P$ exciton.}\label{fig:QR}
\end{figure}

We aim to calculate the exciton eigenfunctions and eigenenergies
for a torus  structure depicted in Fig.
\ref{fig:QR}. 

A cylindrical metallic core is covered by a
semiconductor ring and the height of the structure $H$ (its extension in the growth direction $z$) is of the order of a few monolayers.\\\\
The Hamiltonian for such a system has the form 
\begin{eqnarray}\label{H_disk}
&&H=E_g+\frac{p_{ez}^2}{2m_{e}}+\frac{p_{hz}^2}{2m_{hz}}\nonumber\\
&&-\frac{\hbar^2}{2m_{e}}{\hbox{\boldmath$\nabla$}_e}^{(2D)2}
-\frac{\hbar^2}{2m_{h\parl}}{\hbox{\boldmath$\nabla$}_h}^{(2D)2}\nonumber\\
&&+ V_c+V_{conf e}(\textbf{r}_e)+V_{conf h}(\textbf{r}_h)%\nonumber\\
%&&-\frac{e^2}{4\pi\epsilon_0\epsilon_b\sqrt{(\textbf{r}_e-\textbf{r}_h)^2+(z_e-z_h)^2}},
\end{eqnarray}
where $\textbf{r}=\textbf{i}x+\textbf{j}y$ is a two-dimensional vector, $m_{h\parl}, m_{hz}$ are the in-plane and $z-$direction
hole effective masses, the electron effective mass $m_e$ is
assumed to be isotropic. Due to cylindrical symmetry of the system  we apply the  laplacian in the form 
\begin{equation}
{\hbox{\boldmath$\nabla$}}^{(2D)2}=\frac{1}{r}\frac{\partial}{\partial r}r\frac{\partial}{\partial r}+\frac{1}{r^2}\frac{\partial^2}{\partial\phi^2}+\frac{\partial^2}{\partial z^2}\nonumber.
\end{equation}
The electrons and holes
 in this structure are subjected to the confinement
potentials $V_{conf e,h}$. For simplicity, we consider the hard wall model of such potentials 
\begin{eqnarray}\label{boundarydisk}
&&V_{conf e,h}=V_{conf e,h}(z)+V_{conf \parl}(r_{e,h}),\nonumber\\
& &V_{conf \parl}(r_e)=\left\{ \begin{array}{ll}
0\quad \mbox{for} r_1\leq\quad r_e\leq r_2,\\
\infty\quad \mbox{for}\quad r_e>r_2\quad\hbox{or}\; r_e<r_1,
\end{array}\right.\\
& &V_{conf \parl}(r_h)=\left\{ \begin{array}{ll}
0\quad \mbox{for} r_1\leq\quad r_h\leq r_2,\\
\infty\quad \mbox{for}\quad r_h>r_2\quad\hbox{or}\; r_h<r_1,
\end{array}\right.\nonumber\\
& &V_{conf e}(z_e)=\left\{ \begin{array}{ll}
0\quad \mbox{for}\quad -H/2\leq z_e\leq H/2,\\
 \infty\quad \mbox{for}\quad z_e\geq H/2,\quad\hbox{or}
\quad z_e\leq -H/2,
\end{array}\right.\nonumber\\
& &V_{conf h}(z_h)=\left\{ \begin{array}{ll}
0\quad \mbox{for}\quad  -H/2\leq z_h\leq H/2,\\
\infty\quad \mbox{for}\quad z_h\geq H/2,\quad\hbox{or} \quad
z_h\leq -H/2.
\end{array}\right.\nonumber
\end{eqnarray}
We use the same  calculation scheme as for the
case of spherically symmetric structures, however in the case of the quantum ring 2-dimensional system of cylindrical symmetry imposes specific conditions.\\\\
  The confinement potentials in $z$ direction $V_{conf e,h}(z_{e,h})$ cause that electrons and holes are situated in a standard infinite quantum well, therefor their eigenenergies
$E_{ez}, E_{hz}$ are given by      
%   cause that states $\psi_{ez},\>\psi_{hz}$ of the electrons and holes, which determine the respective eigenvalues $E_{ez}, E_{hz}$, which are given by standard infinite quantum well formula
\begin{equation}\label{eq:qwell}
E_{ez,hz}=\frac{n_{e,h}^2\pi^2a_{e,h}^{*2}}{H^2}R_{e,h}^*.
\end{equation} 
Due to the fact that $H$ reaches the heights of singles monolayers, we will restrict our considerations to the lowest confinement state $N_e=N_h=1$.\\
Like in the case of core-shell structure, we assume that the electron is more massive than the hole and it is located on the metal-dielectric interface, so that $r_e=r_1$ but unlike the spherical symmetry discussed in Appendix A, here the minimum value of $r_e$ cannot be lower than $r_1$ regardless of the structure size. Thus, we put a parameter $\alpha=1$ in Eq. (\ref{Coulomb}). Finally, to include the electron-hole distance in the $z$ direction, the potential is further modified to the form
\begin{equation}\label{culomb1}
V_c^{ring}=\frac{-1}{4\pi\epsilon_0\epsilon_b}\frac{1}{\sqrt{(r-r_1)^2+z^2}},
\end{equation}
where $z=0$ is the center of the structure. The average position of the electron in the z direction is also $z=0$.

We use the same  calculation scheme considering $r$ direction, as for the
case of spherically symmetric structures, but here geometry and the form confinement potentials $V_{conf \parallel}(r_{e,h})$ result in the following forms of Schr\"{o}dinger equations for electrons and holes 
\begin{equation}\label{e1r}
\left[\frac{-\hbar^2}{2m_{e}}{\hbox{\boldmath$\nabla$}_e}^{(2D)2}+V_{conf \parl}(r_e)\right]\psi(r_e)=E_e\psi(r_e),
\end{equation}
\begin{eqnarray}\label{h1}
&&\left[\frac{-\hbar^2}{2m_{h}}{\hbox{\boldmath$\nabla$}_h}^{(2D)2}+V_c^{ring}+V_{conf \parl}(r_h)\right]\psi(r_h)\nonumber\\&&=E_h\psi(r_h).
\end{eqnarray}
Introducing the variable $\tilde{\rho}=\sqrt{E_e}\frac{r_e}{a_e^*}$  one can postulate the  general form of the solution of Eq. (\ref{e1r}) 
\begin{equation}\label{e2}
\psi_e(\tilde{\rho},\phi)=\psi_m(\tilde{\rho})\frac{e^{im\phi}}{\sqrt{2\pi}}.
\end{equation}
Then Eq. (\ref{e1r}) can be transformed into 
\begin{equation}\label{e5} 
\frac{d^2 \psi_m}{d\tilde{\rho}^2}+\frac{1}{\tilde{\rho}}\frac{d\psi_m}{d\tilde{\rho}}+\left(1-\frac{m^2}{\tilde{\rho}^2}\right)\psi_m=0,
\end{equation}
with the solution 
\begin{equation}\label{e6}
\psi_m(\tilde{\rho)}=C_m[Y_m(k\tilde{\rho}) J_m(k\tilde{\rho})+ J_m(k\tilde{\rho})Y_m(k\tilde{\rho})],
\end{equation}
where $J_m, Y_m$ are the Bessel functions of the first and of the second kind, respectively and $k=\sqrt{E_e}$. \\\\
The solution of Eq. (\ref{h1})  for the hole is hydrogen-like eigenfunctions, which radial part $R_h$ satisfy the equation
\begin{eqnarray}\label{Radial_Disk}
&&\frac{\hbar^2}{2m_h}\left[\frac{\partial^2}{\partial r_h^2}+\frac{1}{r_h}\frac{\partial}{\partial r_h}-\frac{m^2}{r_h^2}\right.\nonumber\\&&\left.-V_c^{ring}+V_{conf \parallel}(r_h) \right]R_h=E_h R_h,
\end{eqnarray}

Compared to the spherical core-shell particle, cylindrical symmetry imposes a different set of quantum numbers. As mentioned above, the confinement in the $z$ direction yields electron and hole numbers $n_e=n_h=1$ analogous to a quantum well. In the radial direction, one has  the principal quantum number $n$ and azimuthal quantum number $m$ \cite{Parfitt2002}.

\subsection{Quantum rod}
\begin{figure}[ht!]
\centering
\includegraphics[width=.8\linewidth]{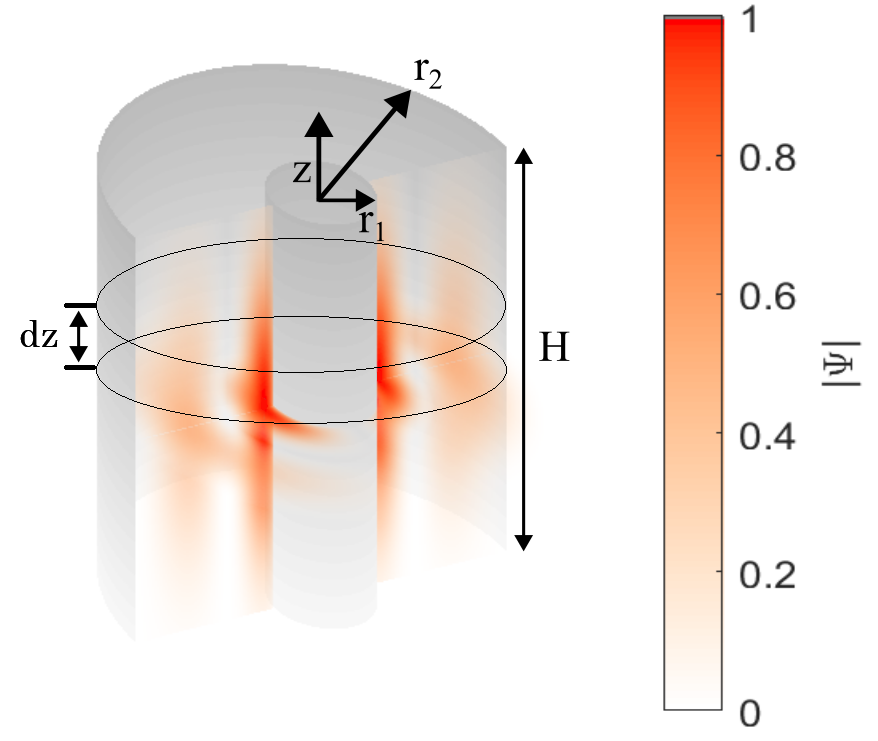}
\caption{Schematic of the quantum rod system; metal cylinder with a radius $r_1$ is surrounded by a layer of Cu$_2$O with radius $r_2$. Color marks horizontal and vertical cross-sections of the hole wavefunction of $2P$ exciton.}\label{fig:QRod}
\end{figure}

 Similar to the quantum ring a cylindrical metallic core of quantum rod is covered by a semiconductor shell, as shown on Fig. \ref{fig:QRod}. 
A quantum rod is an interesting case of a quantum ring system in which the cylinder height $H$ is much greater then this in a quantum ring; it amounts up to several dozen nm and is at least 10 times larger than its diameter. Therefore, one can say that the quantum rod is a structure that is intermediate between 3-dimensional core-shell and 2-dimensional quantum ring.

 The Hamiltonian is given by Eq. (\ref{H_disk}) and the scheme of proceeding is analogous to that presented in the previous sections. The screening and confinement potentials for electrons and holes are the same as in the case of quantum ring, given by Eqs (\ref{culomb1}, \ref{boundarydisk}), however the influence of $H$ is important.\\
In the quantum rod geometry electrons and holes move quickly in
the radial direction, giving rise to a modified, effective Coulomb potential, obtained by averaging with the in-plane wave functions $\psi_m$ of quantum rings corresponding to slices of the quantum rod wih height $dz$ (see Fig. \ref{fig:QRod})
\begin{eqnarray}\label{effpot1}
V_{m}(\zeta)&=&-2\int_{\rho_1}^{\rho_2}\frac{(\psi_{m})^2\rho
d\rho}{\sqrt{\rho^2+\zeta^2}},
\end{eqnarray}
with a relative electron-hole coordinate $$\zeta=\frac{z_e-z_h}{a^*_h}.$$\\
The effective  potential enables one to include the significant size of $H$, which makes a key difference between a quantum ring and quantum rod case. It can be approximated as \cite{NJP}
\begin{equation}
V_m(\zeta)=-\frac{2}{\beta_m+\vert\zeta\vert},
\end{equation}
where
\begin{equation}
\beta_m^{-1}=\beta_N^{-1}=\int_{\rho_1}^{\rho_2}(\psi_{m}(\rho))^2 d\rho.
\end{equation}
The Schr\"{o}dinger equation for electrons and holes can be written in the form
\begin{equation}\label{efEz}
-\frac{d^2}{d\zeta^2}\psi_m-\frac{2}{\beta_m+\vert\zeta\vert}\psi_m=E_{ez,hz}\psi_m.
\end{equation}
Solution of Eq. (\ref{efEz}) gives eigenenergies for electrons and holes in $z$ direction and transverse components of eigenfunctions and eigenenergies can be calculated in the same way as  for quantum ring, as it was presented in the previous subsection (see Eqs. (\ref{H_disk}-\ref{Radial_Disk})).

\section{Calculation results}
For our calculations, we use the material parameters outlined in Table \ref{parametervalues}.
\begin{table}[ht!]
\caption{Band parameter values for Cu$_2$O from \cite{Stolz},
energies in meV, masses in free electron mass $m_0$, lengths in
nm, $\Delta_{LT}$ from \cite{Klingshirn}}
\begin{tabular}{p{.3\linewidth} p{.3\linewidth}p{.3\linewidth}}
\hline
Parameter~ & \hbox{Value}& Ref. \\
\hline $E_g$ & 2172.08&\cite{Stolz}\\
$R^*$&90.88&\cite{Stolz}\\
$R^*_e$&246.6& Eq. (\ref{18})\\
$R^*_h$&144& Eq. (\ref{18})\\
$\Delta_{LTS}$&$5\times 10^{-2}$&\cite{Klingshirn}\\
 $m_e$ & 0.985 &\cite{Stolz}\\
$m_{h}$  &0.575&\cite{Stolz} \\
$\mu$  & 0.363&\cite{Stolz} \\
$\epsilon_b$&7.37&\cite{Stolz}\\
$a^*$&1.07&\cite{Stolz}\\
$a^*_e$&0.396&Eq. (\ref{18})\\
$a^*_h$&0.68&Eq. (\ref{18})\\
\hline\\
\end{tabular} \label{parametervalues}
\end{table}
The exciton Rydberg energies $R_{e,h}^*$ and the excitonic Bohr
radii $a_{e,h}^*$ were obtained by the formulas
\begin{equation}\label{18}
R_{e,h}^*=\frac{m_{e,h}\cdot 13600}{\epsilon_b^2},\quad
a_{e,h}^*=\left(\frac{1}{m_{e,h}}\right)\epsilon_b\cdot 0.0529,
\end{equation}
which are rescaled values obtained for hydrogen atom ($13.6$ eV and 0.0529 nm). Below, we numerically solve the Schr\"{o}dinger equation for the thee specified structures, using the above parameters. The main points of interest are energies of exciton states and the susceptibility spectrum of the Cu$_2$O layer.

\subsection{Exciton energies and spectrum for a core-shell structure}
\begin{figure}[ht!]
a)\includegraphics[width=.95\linewidth]{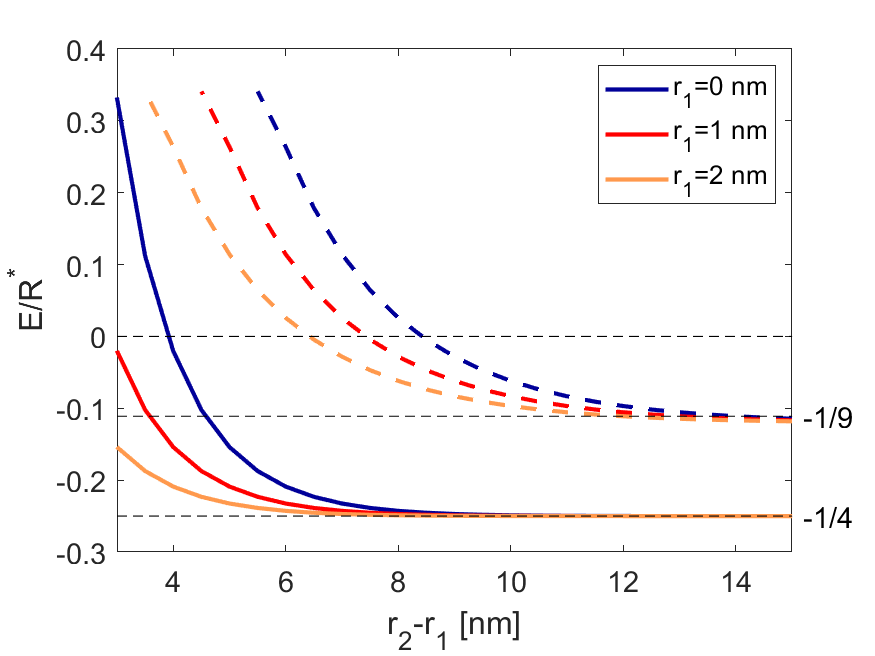}
b)\includegraphics[width=.95\linewidth]{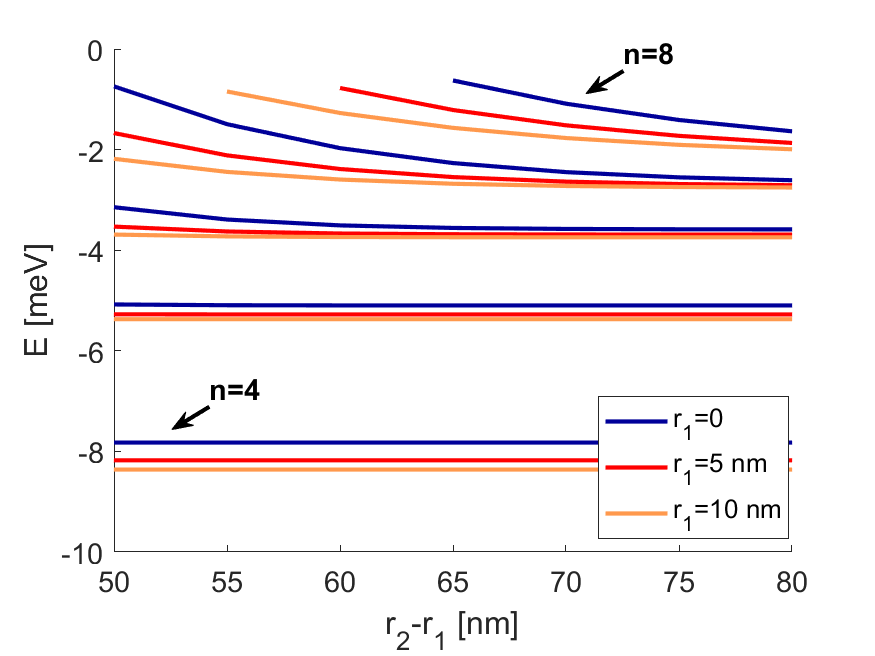}
\caption{a) Energy of a 2$P$ exciton (continuous line) and 3$P$ exciton (dashed line) in a core-shell structure, for three values of $r_1$, as a function of $r_2-r_1$. Asymptotic bulk limits $R^*/4$, $R^*/9$ and $E=0$ are marked by a horizontal, dashed line. b) Energies of n=4-8 excitons in a core-shell structure, calculated for $r_1=0,5,10$ nm.}\label{fig:res1}
\end{figure}

Fig. \ref{fig:res1} a) depicts 2$P$ and 3$P$ exciton energy as a function of the Cu$_2$O layer thickness $r_2-r_1$. The Coulomb potential is taken as $V_c \sim 1/(r-r_1)$. As expected, the confinement potential imposed by finite layer thickness leads to an increase of the exciton energy. In particular, the largest increase of energy is obtained for $r_1=0$ due to the fact that for a given thickness $r_2-r_1$, the total volume of the layer is smallest in the limit $r_1 \rightarrow 0$. Furthermore, one can see that in the case of $r_1=0$, exciton energy reaches 0 then the layer thickness is approximately equal to the exciton radius ($\sim 4$ nm and $\sim 9$ nm for 2$P$ and 3$P$ excitons, correspondingly). Fig. \ref{fig:res1} b) shows the calculation results performed for Rydberg states; due to the large size of high $n$ excitons, the layer thickness $r_2-r_1$ necessary to sustain them is also increased correspondingly. One can notice that the split between energies calculated for various values of $r_1$ becomes more pronounced as the thickness of the layer decreases. The energy of $n=8$ exciton, which is characterized by a radius $\sim 64$ nm, reaches 0 when the layer thickness matches the exciton radius.

\begin{figure}[ht!]
\includegraphics[width=.95\linewidth]{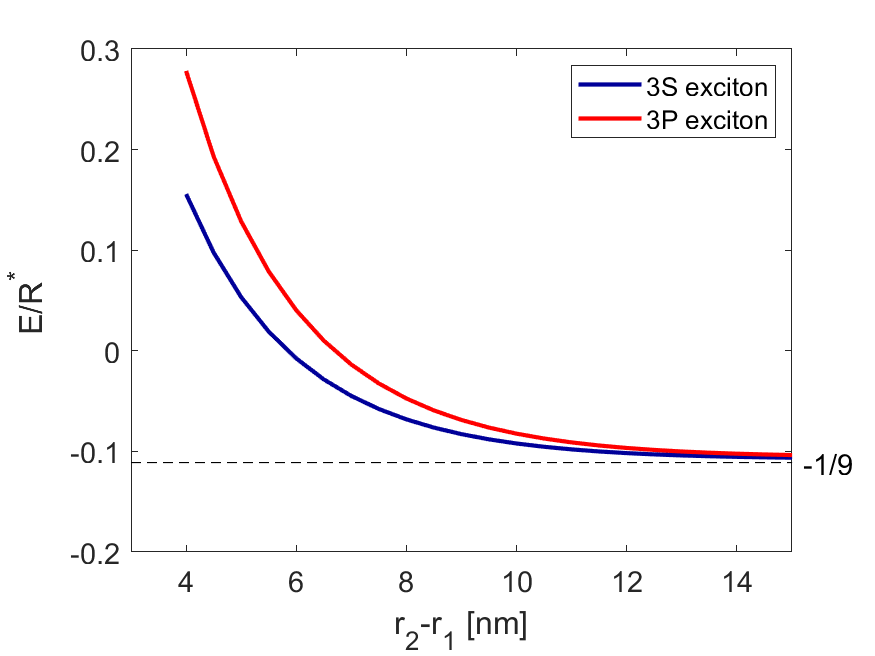}
\caption{Energy of 3$S$ and 3$P$ exciton in a core-shell structure, for $r_1=2$ nm, as a function of $r_2-r_1$. Asymptotic bulk limit $R^*/9$ is marked by a horizontal, dashed line.}\label{fig:res1b}
\end{figure}

One of the consequences of the modified Coulomb potential is the fact that the radial Eq. (\ref{rad_1}) reduces to the form of hydrogenlike equation only in the limit $r_1=0$. Otherwise, the Coulomb part and centrifugal/orbital part of Eq. (\ref{rad_1}) have a different radial dependence. This results in eigenenergies that depend on the quantum number $\ell$. In other words, there is a split between $S$ and $P$ exciton energy. This phenomenon is demonstrated on the Fig. \ref{fig:res1b}.

As expected, the splitting approaches 0 in the bulk limit, matching the behavior of hydrogenlike atom. It should be pointed out that in this paper, we ignore the effects of the complex band structure of Cu$_2$O that lift the degeneracy of the exciton states characterized by the same $n$ \cite{Ertl}. Therefore, in a real excitonic system the effects discussed here will be added on top of the intrinsic $S-P$ energy split of Cu$_2$O excitons. One of the simplifying approaches to include the band structure effects is the introduction of a so-called quantum defect $\delta$ \cite{Kazimierczuk,Ertl}, which yields the exction energy
\begin{equation}\label{eq:defekt}
E=\frac{-R^*}{(n-\delta)^2}.
\end{equation} 
Typically, $\delta$ decreases with increasing $l$ \cite{Ertl}, so that $P$ states have larger energy than $S$ states. Thus, the considered nanostructure increases the intrinsic split.

\begin{figure}[ht!]
\includegraphics[width=.95\linewidth]{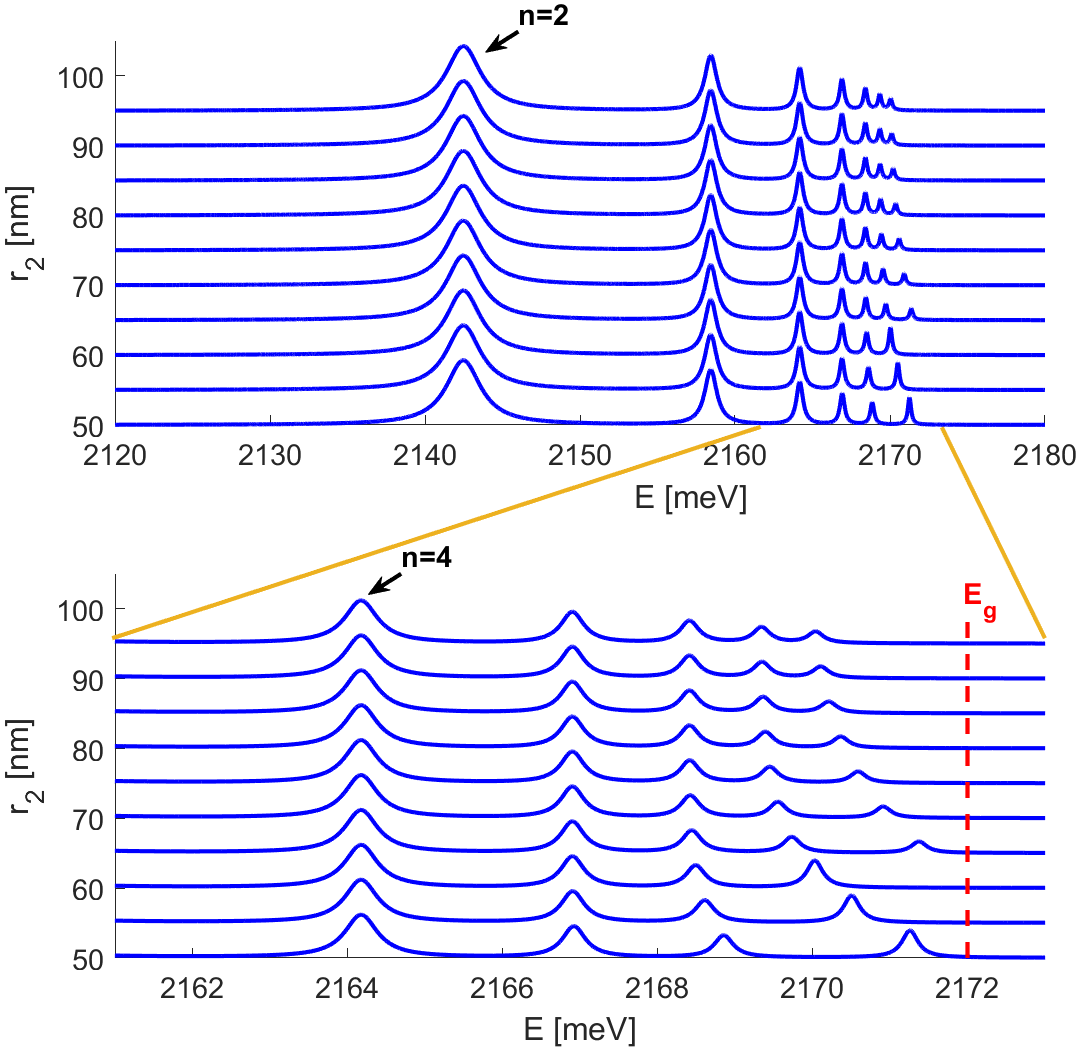}
\caption{Imaginary part of the susceptibility as a function of $r_2$.}\label{fig:res1c}
\end{figure}

The Fig. \ref{fig:res1c} shows the imaginary part of susceptibility calculated from Eq. (\ref{susc}). For clarity, the linewidths are assumed to be equal to bulk values, measured in experiment \cite{Kazimierczuk}. It should be pointed out that excitons in the immediate vicinity of a metal surface will likely exhibit shortened lifetimes due to state hybridization and electron tunneling, as demonstrated for the case of hydrogenlike atoms in \cite{Nord90}. Furthermore, the confinement of exciton in a Cu$_2$O layer leads to increased overlap between electron and hole wavefunction, promoting recombination \cite{Califano}; for a further discussion, see also \cite{Satpathy,Scerri}. In general, the reduction of exciton lifetime (and thus broadening of the excitonic lines) is a highly nonlinear phenomenon dependent on multiple factors, including technical limitations such as surface roughness, residual stress \cite{Hamid_Pillar,Mund2019} on the metal-dielectric boundary etc. For this reason, to limit the scope of this paper, bulk lifetimes are used.
On Fig. \ref{fig:res1c}, the gap energy $E_g \approx 2172$ meV is marked by a red, dashed line. One can see that the spectral lines corresponding to high $n$ states ($n=6,7,8$) move towards higher energy as $r_2$ decreases. In particular, the energy of $8P$ exciton exceeds the gap energy when $r_2$ is smaller than $8P$ exciton radius ($\sim 64$ nm). Therefore, a smaller structure cannot support $8P$ excitons. In general, one can conclude that the structure dimensions put an upper limit on the exciton size (and thus, principal quantum number). 

Another exciton size-related factor influencing the optical spectrum is the so-called Rydberg blockade \cite{Kazimierczuk}; an energy shift caused by exciton-exciton interaction puts an upper limit on the exciton density. In this paper, we assume that the illuminating beam is weak enough that blockade effects are negligible \cite{my_propag}. However, in particularly small nanostructures, the number of excitons that can be sustained simultaneously can be very limited, possibly leading to applications in single photon emitters \cite{Khazali}. 

\subsection{Exciton energies for a quantum ring}
\begin{figure}[ht!]
a)\includegraphics[width=.95\linewidth]{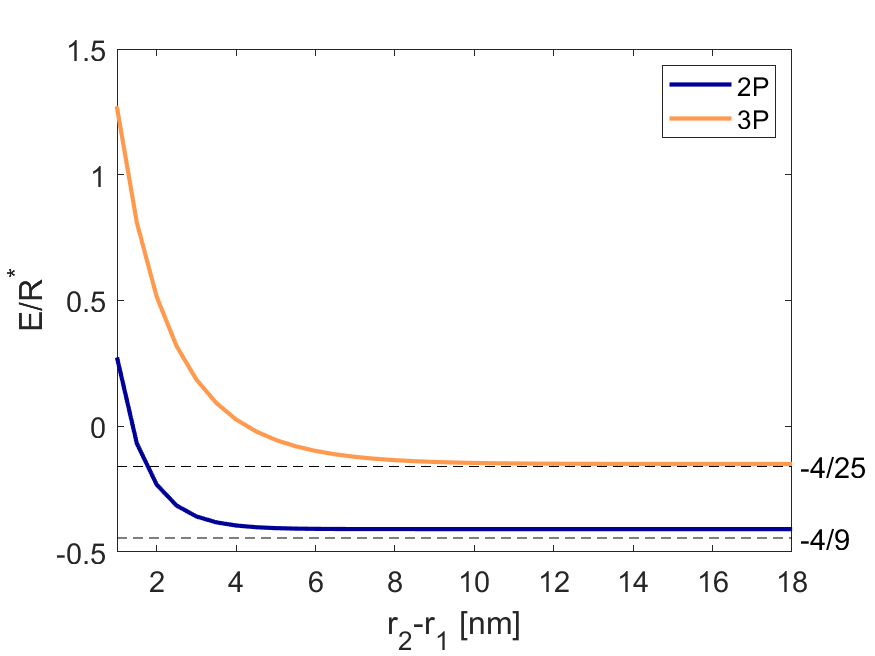}
a)\includegraphics[width=.95\linewidth]{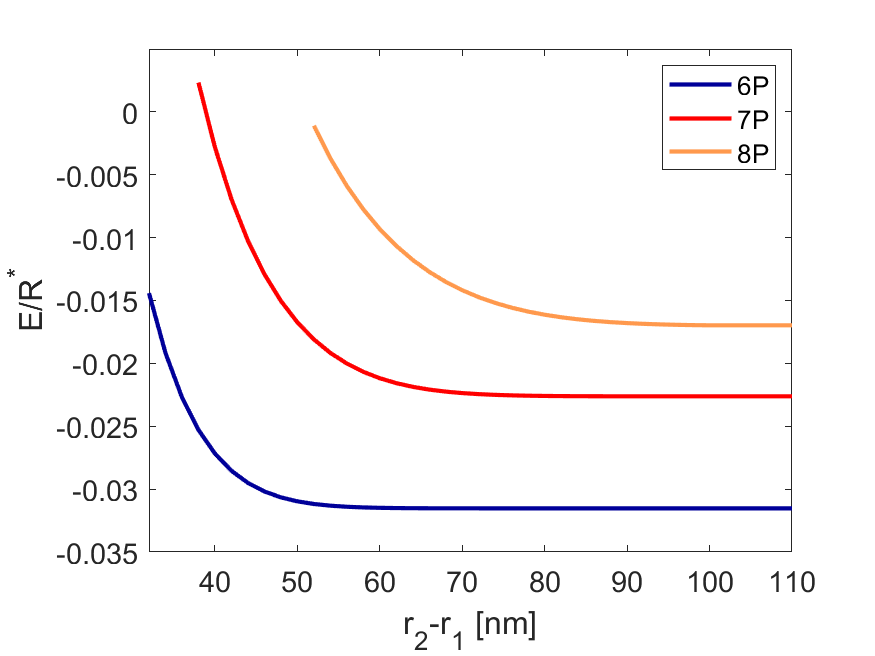}
\caption{Energy of a) 2$P$, 3$P$ and b) $6P,7P,8P$ excitons in a quantum ring with $r_1=2$ nm.}\label{fig:res2}
\end{figure}

In the quantum ring, the exciton is confined to two dimensions, so that the exciton energy in the large radius limit is \cite{Parfitt2002}
\begin{equation}\label{energia_ring}
E=\frac{-R^*}{(n-1/2)^2}.
\end{equation}
Thus, the binding energy is considerably higher than in bulk. This is illustrated on Fig. \ref{fig:res2} a), where energy of 2$P$ and 3$P$ excitons is shown. Energy asymptotes given by Eq. (\ref{energia_ring}) are marked with dashed lines. It should be noted that the confinement energy in the z direction, given by Eq. (\ref{eq:qwell}), is not included here; this energy is constant for a given structure height $H$. The Fig. \ref{fig:res2} b) depicts energy of the Rydberg states; like in the case of core-shell particle, excitons cannot be sustained when their energy becomes positive (so that the excitonic resonance is located above the band gap). However, due to the larger binding energy in the two-dimensional system, slightly stronger confinement is possible than in the core-shell structure. One can see that $8P$ exciton can exist in 50 nm layer, which is less than the radius of bulk $8P$ exciton.

\begin{figure}[ht!]
\includegraphics[width=.95\linewidth]{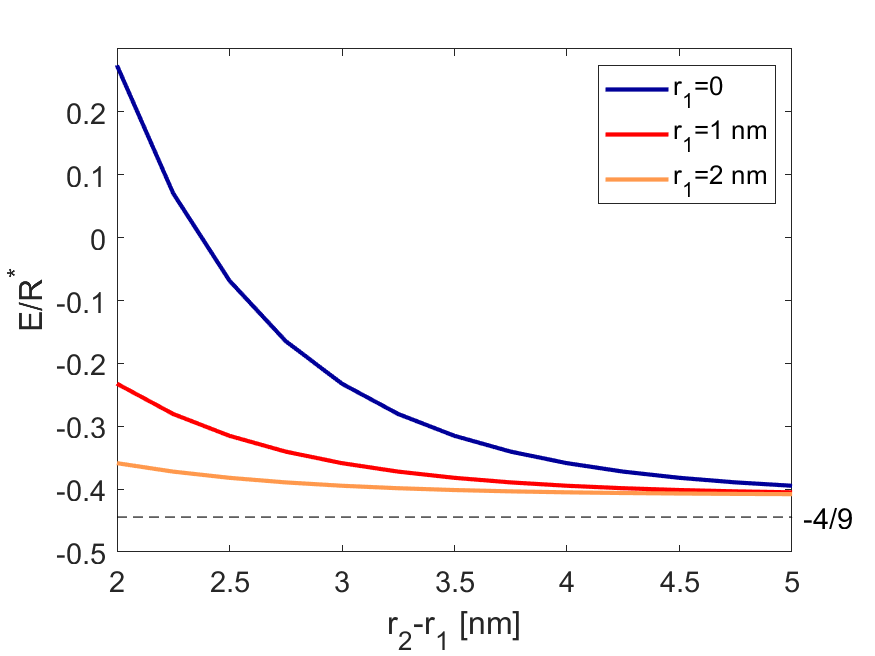}
\caption{Energy of 2P exciton in a quantum ring for various values of $r_1$.}\label{fig:res3}
\end{figure}

\begin{figure}[ht!]
\includegraphics[width=.95\linewidth]{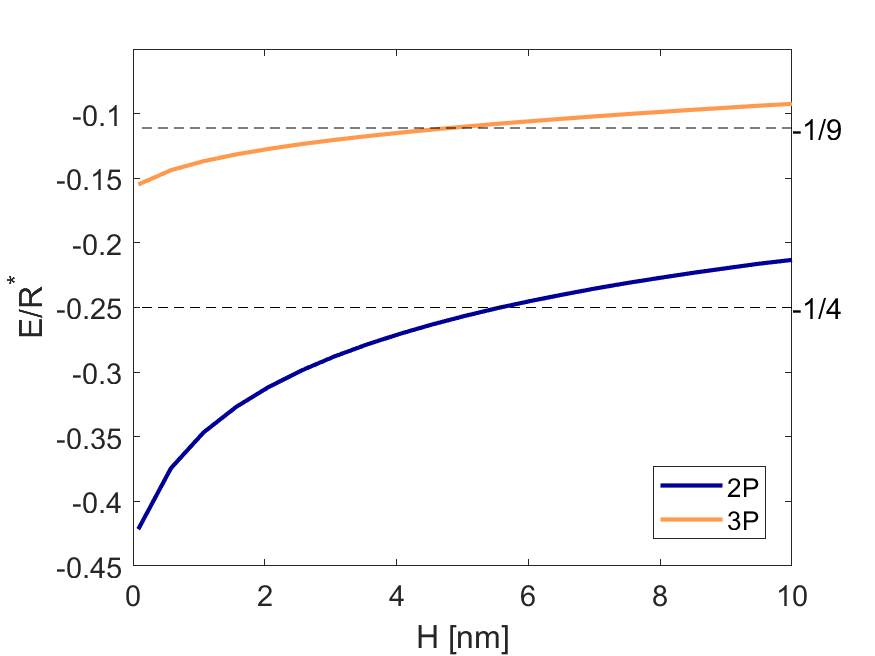}
\caption{Energy of 2$P$ and 3$P$ exciton in a quantum rod, for $r_1=2$ nm, $r_2=10$ nm, as a function of rod height $H$. Asymptotic bulk limits $R^*/n^2$ are marked by a horizontal, dashed lines.}\label{fig:res4a}
\end{figure}

The dependence of energy on $r_1$ exhibits the same behavior as in the spherical symmetry, as demonstrated on Fig. \ref{fig:res3}, where numerically calculated energy for $r_1=0,1,2$ nm is shown. 
Like in the case of core-shell structure, the energy on Fig. \ref{fig:res3} quickly approaches the limit given by Eq. (\ref{energia_ring}) as either the radius of the structure or the thickness of Cu$_2$O layer increases. It should be pointed out that the increased binding energy of the two-dimensional exciton means that the 2$P$ exciton energy shown on Fig. \ref{fig:res3} remains negative even for a very thin Cu$_2$O layer $r_2-r_1 \sim 3$ nm, which is less than the exciton radius.
\\\\
%\subsection{Exciton energies in a quantum rod}
As previously mentioned, the quantum rod is a structure that is intermediate between 3-dimensional core-shell and 2-dimensional quantum ring. Thus, one can expect that the value of the binding energy of exciton in a quantum rod will also be between the two limiting cases. This is shown on Fig. \ref{fig:res4a}.
In the limit of $H \rightarrow 0$, the energy of $2P$ exciton approaches the value of $\frac{-4}{9}R^*$. As discussed in the quantum ring section, the two-dimensional limit is hard to achieve, with the height of the structure being limited to few atomic monolayers. Therefore, as the height $H$ increases, the energy on Fig. \ref{fig:res4a} very quickly approaches the bulk value. Due to the finite $r_2$ and thus radial confinement, the energy eventually exceeds the bulk limit. 

\subsection{Comparison of symmetries}
\begin{figure}[ht!]
\includegraphics[width=.95\linewidth]{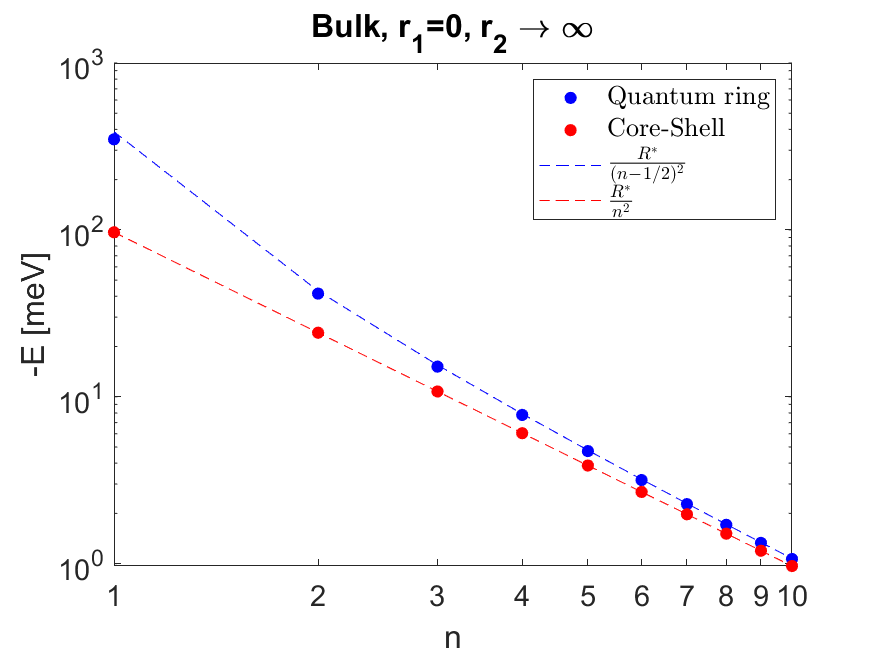}
\caption{Exciton energy as a function of the quantum number n, calculated for core-shell and quantum ring structures.}\label{fig:res4}
\end{figure}

It is interesting to compare some key properties of excitons in the two limiting cases of quantum rings and core-shell particles. First, let's consider the exciton energy in the limit of $r_2 \rightarrow \infty$. As shown on Fig. \ref{fig:res4}, the calculated energies (dots) closely match the theoretical asymptotes (dashed lines) for 3 and 2-dimensional system.
Interestingly, the factor $1/2$ in Eq. (\ref{energia_ring}), describing the quantum ring, can be seen as a special case of quantum defect presented in Eq. (\ref{eq:defekt}). This suggests that more generally, the effects of the dimensionality of the system can be incorporated into quantum defect when a simplified description is needed.

\begin{figure}[ht!]
\includegraphics[width=.95\linewidth]{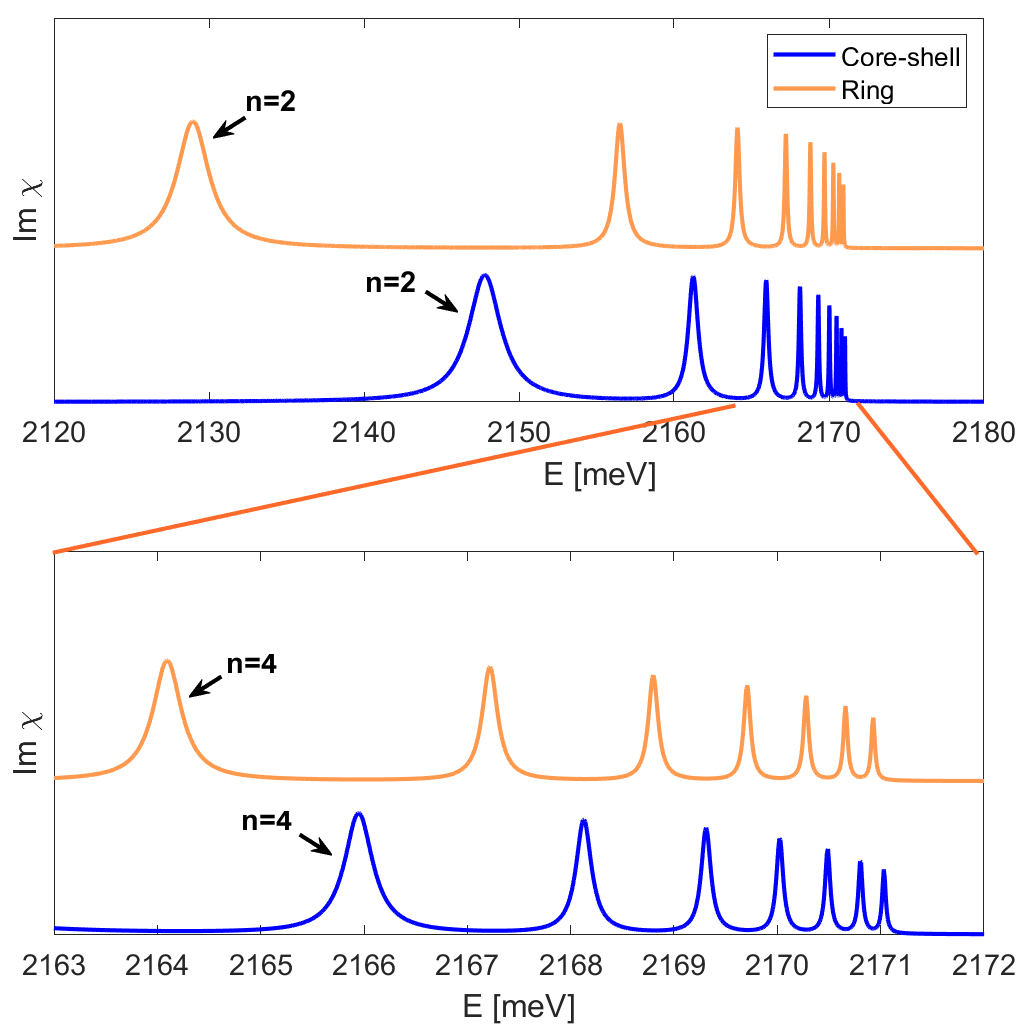}
\caption{Susceptibility spectrum of a core-shell and quantum ring systems, calculated for $r_1=2$ nm and $r_2=100$ nm.}\label{fig:res6}
\end{figure}

Finally, Fig. \ref{fig:res6} shows the imaginary part of susceptibility of the $P$ exciton series calculated from Eq. (\ref{susc}). The calculations were performed in the limit of $r_2 \rightarrow \infty$ and the linewidths $\Gamma$ were taken from bulk measurements \cite{Kazimierczuk}. 
Again, one can observe a large shift of energy spacing between exciton lines when the system transitions from 3 to 2 dimensions. However, both exciton series approach the gap energy $E_g \approx 2172$, which is unaffected by the confinement.

\section{Conclusions}
We have investigated the optical properties of Rydberg excitons in core-shell nanostructures of various dimensionality and symmetry (a core-shell, quantum ring and quantum rod). The dependence of the exciton energy on the geometry of the structure is explored, focusing on the interplay between nanostructure size, exciton size and dimensionality of the system. It is shown that the confinement effects increase the intrinsic energy split between $S$ and $P$ excitons. With the help of Real Density Matrix Approach, we have calculated dielectric susceptibility spectrum of the Cu$_2$O shell, including $P$ exciton series up to $n=10$. The impact of the system geometry and dimensionality on the spectrum is discussed. The results demonstrate that by engineering the composition, thickness and morphology of the semiconductor shell, one can modulate optical properties of such nanostructures to control on demand quantum confinement effects, excitonic energies and positions of excitonic resonances. The results are applicable to both low principal quantum number states as well as Rydberg excitons and pave the way to further studies of exciton-plasmon interactions in the core-shell systems.

\section{Acknowledgment}
Support from National Science Centre, Poland (Project No. OPUS 2025/57/B/ST3/00334), is greatly acknowledged.

\section{Appendix A: effective Coulomb potential}
\begin{figure}[ht!]
\includegraphics[width=.95\linewidth]{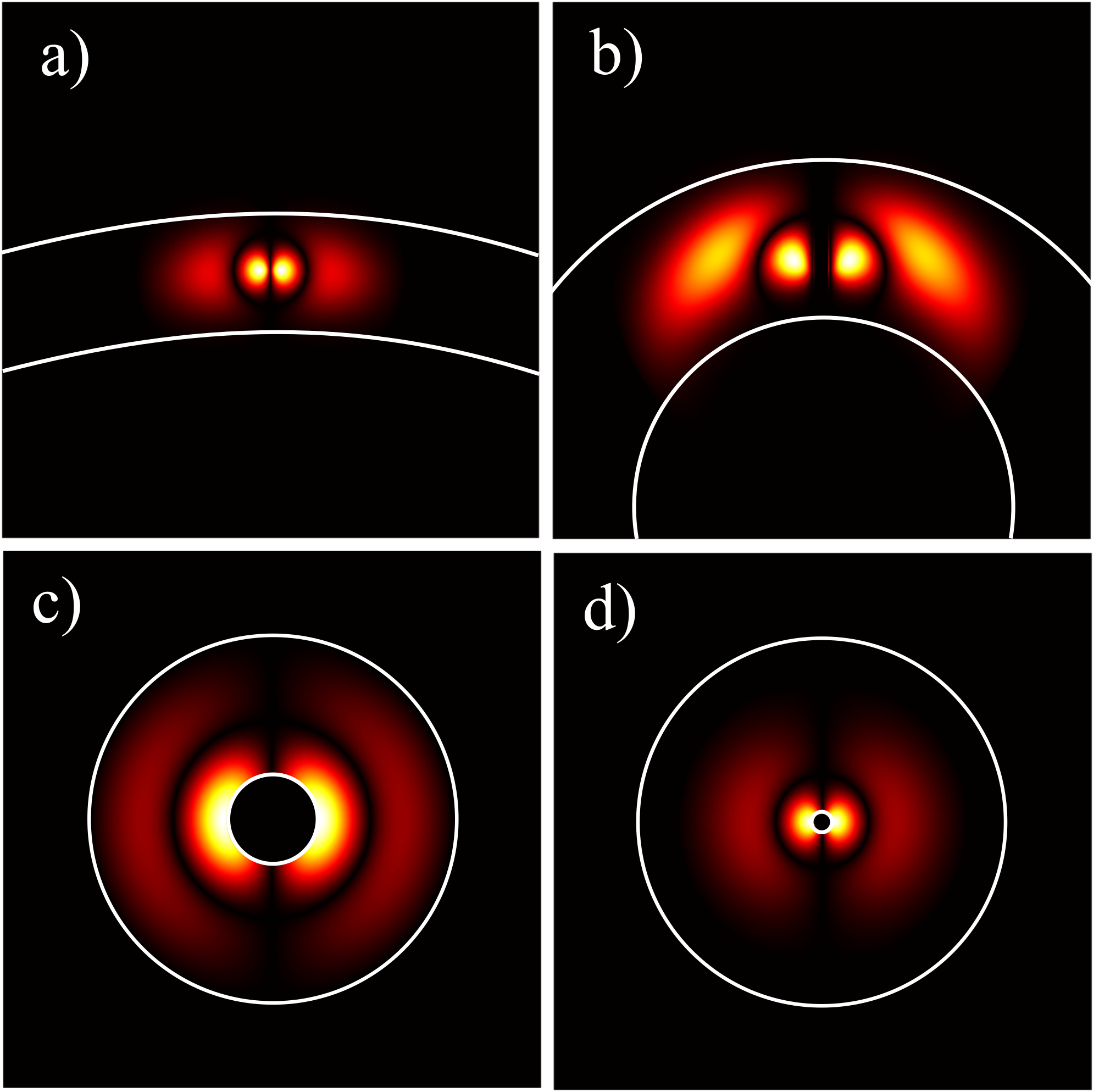}
\caption{Two-dimensional cross-section of the probability density of hole wavefunction of 2P exciton in core-shell structure for a) $r_1>>r_{exc}$, $\alpha=0$ b) $r_1 \sim r_{exc}$, $\alpha>0$, c) $r_1 < r_{exc}$, $\alpha=1$, d) $r_1 << r_{exc}$, $\alpha=1$. White lines depict inner and outer boundary of the Cu$_2$O layer.}\label{fig:ralpha}
\end{figure}

The structures considered in this paper are characterized by a metal central core with radius $r_1$ that acts as an infinite potential barriers for electrons and holes. Thus, both charge carriers are confined to the Cu$_2$O shell (with thickness $r_2-r_1$) surrounding the core, with their wavefunctions having considerable overlap \cite{Califano}. In other words, the average exciton-hole distance is reduced compared to the bulk medium. Let's first consider the spherical core-shell nanoparticle. The electron is assumed to be considerably heavier than the hole. To preserve system symmetry, a modified radial Coulomb potential can be employed. One can distinguish three limiting cases:
\begin{enumerate}
\item When $r_1 \rightarrow 0$ (or, more generally, $r_1$ being much smaller than average exciton radius), one can use the standard Coulomb potential $V_c \sim 1/r$ for the hole. The electron wavefunction is assumed to be evenly distributed around the central core, which leads to the above-mentioned charged sphere potential.
\item When $r_1 \rightarrow \infty$ (or just much larger than exciton radius), the curvature of the system becomes irrelevant and the problem is reduced to one-dimensional quantum well. In the plane of the well, the potential is again $V_c \sim 1/r$.
\item In the intermediate regime, the minimal distance between electron and structure center is limited by the core, with limiting case $V_c \sim 1/(r-r_1)$.
\end{enumerate}
Schematic representation of the exciton wavefunctions for the above cases is shown on Fig. \ref{fig:ralpha}.
To include all of the above cases, we use a potential
\begin{equation}\label{gen_coulomb}
V_c=\frac{-1}{4\pi\epsilon_0\epsilon_b}\frac{1}{r-\alpha r_1},
\end{equation}
with some parameter $0 \leq \alpha \leq \frac{r_2}{r_1}$. The above potential has either radial or cylindrical symmetry, matching the discussed nanostructure (and its boundary conditions). The value $\alpha=1$ means that the electron is located on the core surface and $\alpha>1$ represents electron placed above it. 

It should be stressed that the proposed potential is an approximation that allows for both semi-analytical and straightforward numerical solutions; as discussed in \cite{Kovalenko}, the presence of a planar potential barrier (corresponding approximately to Fig. \ref{fig:ralpha} a) and to some extent b)) breaks the spherical symmetry of the Coulomb potential. Thus, $\alpha$ should be treated as a free parameter that allows for inclusion of nonsymmetric potential and maintains continuity between solutions in various regimes.

\section{Appendix B: numerical solution of Schr\"{o}dinger equation}
In this manuscript, analytical solutions have been supplemented by numerical calculations based on the solution of one or multiple one-dimensional Schr\"{o}dinger equations. Let's consider the Eq. (\ref{rad_1}) for the radial wavefunction of the hole. It can be put in form
\begin{equation}\label{num1}
\frac{-\hbar^2}{2m_h}\frac{d^2 R}{dr^2}+V'R=ER,
\end{equation}
where 
\begin{equation}
V'(r)=\frac{e^2}{4\pi\epsilon_0\epsilon_b r}-\frac{\hbar^2}{2m_h}\frac{\ell(\ell+1)}{r^2}+V(r)
\end{equation}
and $V(r)$ is the confinement potential. In the numerical calculations, infinite potential wall is approximated by very large but finite energy
\begin{equation}
V(r<r_1)=V(r>r_2)=100R^*,
\end{equation}
where $R^*$ is the Rydberg energy. In order to solve Eq. (\ref{num1}), a finite spatial step $\Delta r=0.01$ nm is introduced. Thus, the radius $r$ can only take discrete values denoted $r_n=n\Delta r$. The range of possible values of $r$ is set to $0\leq r \leq 2r_2$, so that it extends beyond the considered structure.

The spatial derivative can be approximated by using the first term of Taylor series to obtain
\begin{eqnarray}
\frac{\partial f}{\partial r}(r_{n+\frac{1}{2}}) \approx \frac{f(r_{n+1})-f(r_n)}{\Delta r},\nonumber\\
\frac{\partial f}{\partial r}(r_{n-\frac{1}{2}}) \approx \frac{f(r_{n})-f(r_n-1)}{\Delta r},
\end{eqnarray}
where the lower index $\frac{1}{2}$ expresses the fact that the difference between two discrete values of the function represents the derivative between these values (a so-called central difference approach \cite{Dijk}). Higher order terms $\sim \Delta r^2$, $\Delta r^3$.. are discarded. The second derivative is
\begin{eqnarray}
&&\frac{\partial^2 f}{\partial r^2}(r_n) \approx \frac{\frac{\partial f}{\partial r}(r_{n+\frac{1}{2}})-\frac{\partial f}{\partial r}(r_{n-\frac{1}{2}})}{\Delta r}\nonumber\\
&&=\frac{f(r_{n+1})-2f(r_n)+f(r_n-1)}{\Delta r^2}.
\end{eqnarray}
Thus, the discrete representation of Eq. (\ref{num1}) is
\begin{equation}\label{num2}
R_{n+1}=2R_n-R_{n-1}-\frac{2m_h}{\hbar^2}\frac{1}{\Delta r^2}(E-V')R_n.
\end{equation}
The above relation allows one to calculate the next value $R_{n+1}$ basing on the two previous values $R_{n}$ and $R_{n-1}$. The two first values $R_0$, $R_1$ are initially set to $[1,1]$ or $[0,1]$ depending on the expected parity of the wavefunction. Due to the relatively small $\Delta r=0.01$ nm, the above first-order scheme provides sufficient accuracy and stability for the problems considered in this manuscript; possible refinements include higher order Runge-Kutta method or Numerov's algorithm \cite{Caruso, Zlotnik}.

\begin{figure}[ht!]
\centering
\includegraphics[width=.9\linewidth]{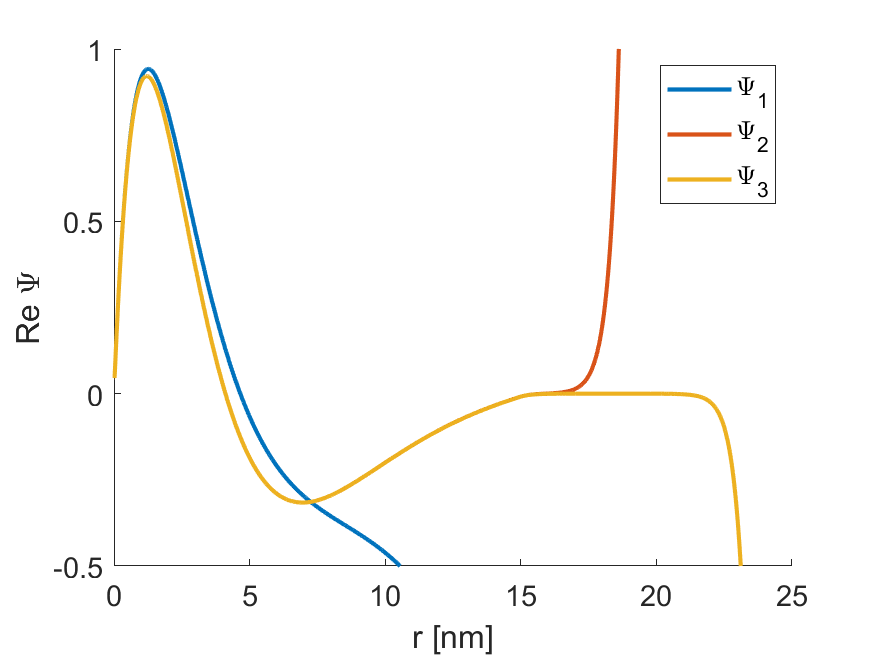}
\caption{Real part of the radial wavefunction of 2S exciton; three consecutive approximations are shown.}\label{fig:num}
\end{figure}

In order to solve the Eq. (\ref{num2}), the energy $E$ has to be known. In order to find this energy, one can use some first estimate $E_0$ and then construct successive, better estimates $E_1$, $E_2$... through the method of bisection. Specifically, a key characteristic of the Eq. (\ref{num2}) is that the calculated function $R(r)$ will exactly match the assumed boundary conditions only when the energy also matches one of the eigenenergies. If the guess $E_0$ is incorrect, the function $R(r)$ will be nonzero for $r=r_2$ and, more generally, diverge when $r \rightarrow \infty$. By using the sign of the function at the point of divergence, one can find out whether the energy is over- or underestimated. The bisection algorithm is as follows:
\begin{enumerate}
\item Define the numerical grid $r_n=0,1...N\Delta r$ and potential $V'(r_n)$.
\item Assume initial energy guess $E_0$ and energy step $\Delta E$.
\item Use $E_0$ to calculate first wavefunction approximation $R_0$.
\item Update energy $E_1=E_0+\Delta E$.
\item Use $E_1$ to calculate $R_1$.
\item If $R_0(r_N)R_1(r_N)<0$, then update energy step $\Delta E'=-\Delta E/2$.
\item Repeat steps 4-6 until sufficient accuracy is obtained.
\end{enumerate}
The process is illustrated on Fig. \ref{fig:num}, where three consecutive steps of the calculation are presented. The resulting in wavefunction approximations diverge to $\pm \infty$, with better approximations diverging later.


\begin{thebibliography}{99}


\bibitem {Kazimierczuk}
T. Kazimierczuk, D. Fr\"{o}hlich, S. Scheel, H. Stolz and M. Bayer, \emph{Giant Rydberg Excitons in Cuprous Oxide}, Nature \textbf{514}, 344 (2014).

\bibitem{Heck017}
J. Heck\"{o}tter, M. Freitag, D. Fr\"{o}hlich, M. Assmann, M. Bayer, M. A. Semina and M. M. Glazov, \emph{High-resolution study of the yellow excitons in Cu2O subject to an electric field}, Phys. Rev. B \textbf{95}, 035210 (2017).

\bibitem{Schweiner2017}
F. Schweiner, J. Ertl, J. Main,  G. Wunner and Ch. Uihlein, \emph{Exciton-polaritons in cuprous oxide: Theory and comparison with experiment}, Phys Rev. B 96, 245202 (2017).

\bibitem{Thomas}
P.Chakrabarti, K.Morin, D.Lagarde, X.Marie and T.Boulier, \emph{Direct measurement of the lifetime and coherence time of Cu$_2$O Rydberg excitons}, Phys. Rev. Lett. \textbf{134},
126902 (2025).

\bibitem{Ertl}
J. Ertl, P. Rommel and J. Main, \emph{Quantum defects of Rydberg excitons in cuprous oxide: a semiclassical spherical model}, Eur. Phys. J. Spec. Top. (2025), \url{https://doi.org/10.1140/epjs/s11734-025-02063-3}

\bibitem{Kim}
N.Y. Kim, \emph{Rydberg excitons in Cu$_2$O: exaggereted quantum optical properties in emerging quantum phenomena}, Journal of the Optical Society of America B \textbf{43}, A2278 (2026).

\bibitem{Heck2017}
J. Heck\"{o}tter, M. Freitag, D. Fr\"{o}hlich, M. Assmann, M. Bayer, M. A. Semina and M. M. Glazov, \emph{Scaling laws of Rydberg excitons}, Phys. Rev. B \textbf{96}, 125142 (2017).

\bibitem{Steinhauer}
S. Steinhauer, M. Versteegh, S. Gyger, A. Elshaari, B. Kunert, A. Mysyrowicz and V. Zwiller, \emph{Rydberg excitons in Cu2O microcrystals grown on a silicon platform}, Nature Communications Materials \textbf{1}, 11, 2 (2020).

\bibitem{Morin2022}
C. Morin, J. Tignon, J. Mangeney, S. Dhillon, G. Czajkowski, K. Karpiński, S. Zielińska-Raczyńska, D. Ziemkiewicz and T. Boulier, \emph{Self-Kerr effect across the yellow Rydberg series of excitons in Cu2O}, Phys. Rev. Lett. \textbf{129}, 137401 (2022).

\bibitem{Sylwiaqb}
S.Zieli\'{n}ska-Raczy\'{n}ska and D. Ziemkiewicz, \emph{Quantum interference of Rydberg excitons in Cu$_2$0: Quantum beats}, \emph{Phys. Rev. B} \textbf{111}, 205201 (2025).

\bibitem{Neubauer}
A. Neubauer, J. Heck\"{o}tter, M. Ubl, M. Hentschel, B. Panda, M. Assmann, M. Bayer and H. Giessen, \emph{Spectroscopy of nanoantenna-covered Cu2O: Towards enhancing quadrupole
transitions in Rydberg excitons}, Phys. Rev. B \textbf{106}, 165305 (2022).

\bibitem{Hamid}
K. Orfanakis, S. Rajendran, H. Ohadi, S. Zielińska-Raczyńska, G. Czajkowski, K. Karpiński and D. Ziemkiewicz, \emph{Quantum confined Rydberg excitons in Cu$_2$O nanoparticles}, Phys. Rev. B \textbf{103}, 245426 (2021). 

\bibitem{Qdisk}
D. Ziemkiewicz, G. Czajkowski, K. Karpinski and S. Zielinska-Raczynska, \emph{Excitons in Cu$_2$O: From quantum dots to bulk crystals and additional boundary conditions for Rydberg exciton-polaritons}, Phys. Rev. B \textbf{101}, 205202 (2020).

\bibitem{Belov2024}
P. A. Belov, F. Morawetz, S. O. Krüger, N. Scheuler, P. Rommel, J. Main, H. Giessen and S. Scheel, \emph{Energy states of Rydberg excitons in finite crystals: From weak to strong confinement}, Phys. Rev. B \textbf{109}, 235404 (2024).

\bibitem{Konzel2020}
A. Konzelmann, B. Frank and H. Giessen, \emph{Quantum confined Rydberg excitons in
reduced dimensions}, J. Phys. B: At. Mol. Opt. Phys. \textbf{53}, 024001 (2020).

\bibitem{Takahata2018}
M. Takahata, K. Tanaka and N. Naka, \emph{Nonlocal optical response of weakly confined excitons in Cu2O mesoscopic films}, Phys. Rev. B \textbf{97}, 205305 (2018).

\bibitem{Sharma}
S. Sharma, P. Phogat, J. Thakur and N. L. Singh, \emph{Core-Shell Nanomaterials From Fundamentals to Applications}, 2026, Wiley-VCH GmbH.

\bibitem{Bartolucci2024}
S. F. Bartolucci, A. C. Leff, and J. A. Maurer, \emph{Gold-copper oxide core-shell plasmonic
nanoparticles: the effect of pH on shell stability and mechanistic insights into shell formation}, Nanoscale Adv. \textbf{6}, 2499 (2024).

\bibitem{Meir}
N. Meir, I. Plante, K. Flomin, E. Chockler, B. Moshofsky, M. Diab, M. Volokh and T. Mokar, \emph{Studying the chemical, optical and catalytic properties of noble metal (Pt, Pd, Ag, Au)-Cu2O core-shell nanostructures grown via a general approach}, J. Mater. Chem. , \textbf{1}, 1763 (2013).

\bibitem{Wang}
H. Wang, J. Lin, H. Zhang, Y. Zhang, and J. Li, \emph{Plasmonic Core-Shell Materials: Synthesis, Spectroscopic Characterization, and Photocatalytic Applications}, Acc. Mater. Res. \textbf{3}, 187-198 (2022).

\bibitem{Rana2021}
S. Rana, S. Kabi, K. Misra and S. Chattopadhyay, \emph{Exciton and Bi-exciton Binding Energy Calculation in a Core Shell Quantum Dot}, IOP Conf. Series: Materials Science and Engineering \textbf{1080}  012012 (2021).

\bibitem{Hibibi}
M. Hbibi, O. Mommadi, S. Chouef, R. Boussetta, L. Belamkadem, A. El Moussaouy, F. Falyouni, C. M. Duque, J. A. Vinasco and C. A. Duque, \emph{Finite confinement potentials, core and shell size effects on excitonic and electron‑atom properties in cylindrical core/shell/shell quantum dots}, Scientific Reports \textbf{12}, 14854 (2022).

\bibitem{Nandan}
Y. Nandan and M. S. Mehata, \emph{Wavefunction Engineering of Type-I/Type-II Excitons of CdSe/CdS Core-Shell Quantum Dots}, Scientific Reports \textbf{9}:2, 1 (2019). 

\bibitem{Antos2014}
T.J. Antosiewicz, S. P. Apell and T. Shegai, \emph{Plasmon-Exciton Interactions in a Core-Shell Geometry: From Enhanced Absorption to Strong Coupling},  ACS Photonics \textbf{1}, 454 (2014).

\bibitem{Stete}
F.Stete, W.Koopman, C.Henkel, O.Benson, G.Kewes and M.Barghee, \emph{Optical Spectra of Plasmon-Exciton Core-Shell Nanoparticles: A Heuristic Quantum Approach}, ACS Photonics \textbf{10}, 2511 (2023).

\bibitem{Dima}
D. Ziemkiewicz, D. Knez, E. Garcia, S. Zielińska-Raczyńska, G. Czajkowski, A. Salandrino, S. Kharnitsev, A. Noskov, E. Potma and D. Fishman, \emph{Two-photon absorption in silicon using the real density matrix approach}, J. Chem. Phys. \textbf{161}, 144117 (2024).

\bibitem{Abramowitz}
M. Abramowitz and I. Stegun, eds.  Handbook of Mathematical
Functions.  Dover Publications, (ISBN 0-486-61272-4) New York,
USA, 1964,. http://www.ams.org/notices/201107/rtx110700905p.pdf.


\bibitem{Parfitt2002}
D. G. W. Parfitt and M. E. Portnoi, \emph{The two-dimensional hydrogen atom revisited}, J. Math. Phys., \textbf{43}, 10, 4681 (2002).

\bibitem{NJP}
S. Zielińska-Raczyńska, D.A. Fishman, C. Faugeras, M. Potemski,
P. Loosdrecht, K. Karpiński, G. Czajkowski and D. Ziemkiewicz, \emph{Magneto-excitons in Cu2O: theoretical model from weak to high
magnetic fields}, New J. Phys. \textbf{21}, 103012, (2019).


\bibitem{Stolz}
H. Stolz, F.Sch\"{o}ne and D.Semkat, \emph{Interaction of Rydberg excitons in cuprous oxide with phonons and photons: optical linewidth and polariton effect}, New J. Phys. \textbf{20}, 023019 (2018).

\bibitem{Klingshirn}
C. Klingshirn, \emph{Semiconductor Optics}, 2012 Berlin, Springer.

\bibitem{Nord90}
P. Nordlander and J.C. Tully, \emph{Energy shifts and broadening of atomic levels near metal surfaces}, Phys. Rev. B \textbf{42}, 9, 5564 (1990).

\bibitem{Califano}
M. Califano, \emph{Disentangling the role of wave function overlap, exciton-exciton interaction, carrier confinement and shape in the properties of spherical quantum wells}, Nano Research \textbf{18(2)}, 94907162 (2025).

\bibitem{Satpathy}
S. Satpathy, \emph{Eigenstates of Wannier excitons near a semiconductor surface}, Phys. Rev. B \textbf{28}, 4585 (1983).

\bibitem{Scerri}
D. Scerri, T. S. Santana, B. D. Gerardot and E. M. Gauger, \emph{Method of images applied to driven solid-state emitters}, Phys. Rev. B \textbf{95}, 165403 (2017).

\bibitem{Hamid_Pillar}
A. S. Paul, S. K. Rajendran, D. Ziemkiewicz, T. Volz and H. Ohadi, \emph{Local tuning of Rydberg exciton energies in nanofabricated Cu2O pillars}, Communications Materials \textbf{5}, 43 (2024).

\bibitem{Mund2019}
J. Mund, C. Uihlein, D. Fr\"{o}hlich, D. R. Yakovlev and M. Bayer, \emph{Second harmonic generation on the yellow 1S exciton in Cu2O in symmetry-forbidden geometries}, Phys. Rev. B \textbf{99}, 195204 (2019).

\bibitem{my_propag}
S. Zielińska-Raczyńska and D. Ziemkiewicz, \emph{Propagation of short pulses in Rydberg exciton medium under blockade conditions}, Phys. Rev. B \textbf{113}, 245307 (2026).

\bibitem{Khazali}
M. Khazali, K. Heshami and C. Simon, \emph{Single-photon source based on Rydberg exciton blockade}, J. Phys. B: At. Mol. Opt. Phys. \textbf{50}, 215301 (2017).

\bibitem{Kovalenko}
A. F. Kovalenko, and M. F. Holovko, \emph{A hydrogen-like atom near a potential barrier}, J. Phys. E: At. Mol. Opt. phys. \textbf{25} 233 (1992). 

\bibitem{Dijk}
W. Dijk, \emph{On numerical solutions of the time-dependent Schrödinger equation}, Am. J. Phys. \textbf{91}, 826 (2023).

\bibitem{Caruso}
F. Caruso, V. Oguri and F. Silveira, \emph{Applications of the Numerov method to simple quantum
systems using Python}, Revista Brasileira de Ensino de Física \textbf{44}, e20220098 (2022).

\bibitem{Zlotnik}
A. Zlotnik and O. Kireeva, \emph{On compact 4th order finite-difference schemes for the wave equation}, Mathematical Modeling and Analysis \textbf{26(3)}, 479 (2021).

\end{thebibliography}
\end{document}